# Polarization modulation instability in a nonlinear fiber Kerr resonator

Julien Fatome[1,2,3*], Bertrand Kibler[1], François Leo[4], Abdelkrim Bendahmane[1], Gian-Luca Oppo[5], Bruno Garbin[2,3,6], Stuart G. Murdoch[2,3], Miro Erkintalo[2,3] and Stéphane Coen[2,3]

[1]*Laboratoire Interdisciplinaire Carnot de Bourgogne, UMR 6303 CNRS Université Bourgogne Franche-Comté, Dijon, France*
[2]*Department of Physics, The University of Auckland, Private Bag 92019, Auckland 1142, New Zealand*
[3]*The Dodd-Walls Centre for Photonic and Quantum Technologies, New Zealand*
[4]*OPERA-Photonique, Université Libre de Bruxelles, CP 194/5, 50 Av. F. D. Roosevelt, B-1050 Brussels, Belgium*
[5]*SUPA and Department of Physics, University of Strathclyde, Glasgow G4 0NG, Scotland, European Union*
[6]*Université Paris-Saclay, CNRS, Centre de Nanosciences et de Nanotechnologies, 91120, Palaiseau, France*

*Corresponding author: Julien.Fatome@u-bourgogne.fr*

**We report on the experimental and numerical observation of polarization modulation instability (PMI) in a nonlinear fiber Kerr resonator. This phenomenon is phased-matched through the relative phase detuning between the intracavity fields associated with the two principal polarization modes of the cavity. Our experimental investigation is based on a 12-m long fiber ring resonator in which a polarization controller is inserted to finely control the level of intra-cavity birefringence. Depending on the amount of birefringence, the temporal patterns generated via PMI are found to be either stationary or to exhibit a period-doubled dynamics. Experimental results are in good agreement with numerical simulations based on an Ikeda map for the two orthogonally polarized modes. Our study provides new insights into the control of modulation instability in multimode Kerr resonators.**

Modulation instability (MI) is a nonlinear phenomenon characterized by the exponential growth and evolution of periodic perturbations on top of an intense continuous-wave (cw) laser beam [1, 2]. Underpinned by a nonlinearly phase-matched parametric process, it is associated with a transfer of energy from a narrow pump frequency component to a pair of sidebands arranged symmetrically around the pump. In single-pass optical fiber propagation, MI can be naturally phase-matched through a balance between anomalous group-velocity dispersion and Kerr nonlinearity [1, 2]. In contrast, more general phase-matching conditions are possible in the context of passive Kerr resonators, such as fiber ring cavities, because of the crucial role played by the systems' boundary conditions [3–5]. Various configurations of MI have been investigated in that context, including MI in the normal dispersion regime, MI via bichromatic or incoherent driving, as well as competition between MI and Faraday or period-doubled (P2) instabilities [6–12]. Moreover, at variance with single-pass propagation, MI in Kerr resonators can lead to the emergence of *stationary* periodic (Turing) patterns; such patterns are now understood to be intimately related to temporal cavity solitons and microresonator optical frequency combs [13–16].

Birefringence, and nonlinear coupling between the polarization components of light, is also known to contribute to the phase-matching of parametric processes. This leads to polarization MI (PMI) and the emergence of vector temporal patterns [17-19]. In driven resonators, PMI has only been investigated theoretically so far [20–22], but recent demonstrations of orthogonally-polarized dual comb generation in microresonators are sparking a renewed interest in this process [23]. In this Letter, we report on the direct experimental observation of PMI in a passive Kerr resonator. Our experimental test-bed is based on a normally dispersive fiber ring cavity that incorporates a polarization controller for adjustment of the intra-cavity birefringence. This localized birefringence gives rise to a relative phase detuning between the two orthogonal polarization modes of the cavity, which in turn affects the frequency shift of the PMI sidebands. We also find that birefringence can lead to period-doubled (P2) dynamics, characterized by a two round-trip cycle. Our experimental results are in good agreement with theoretical predictions and numerical simulations based on an iterative two-component Ikeda map.

The experimental setup is displayed in Fig. 1(a). It consists of a $L$ = 12-m long passive fiber ring cavity with a finesse $F$ of about 27, mainly built out of spun fiber. To avoid competition with scalar MI [2], we have built a cavity with normal group-velocity dispersion estimated to $\beta_2$ = 47 ps$^2$/km, a value large enough to neglect third-order dispersion. Also, the use of a *spun* fiber (nearly isotropic) avoids group-velocity mismatch between the polarization components. Additionally, to prevent any additional source of bending-induced birefringence, the fiber was carefully off-spooled and wound directly on our experimental board with a large 50-cm diameter. We estimate that this causes a birefringence Δ$n$ no greater than 10$^{-8}$ [24], which can be neglected in our study.

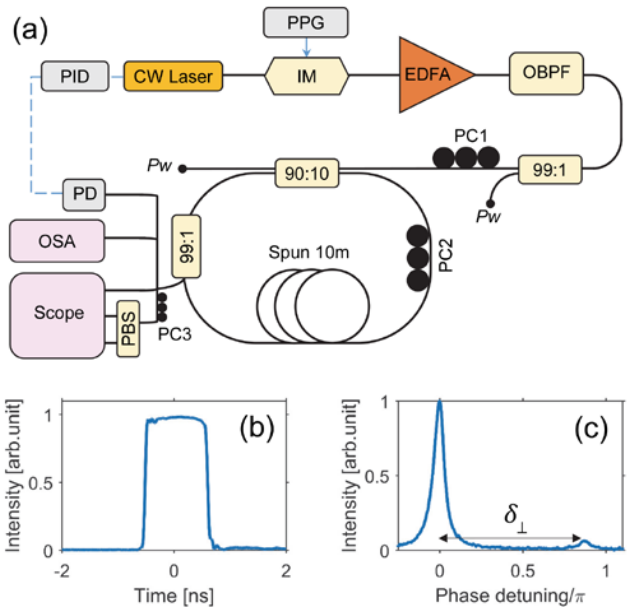

**Fig. 1.** (a) Experimental setup. PPG: pulse pattern generator, IM: intensity modulator, PC: polarization controller, EDFA: erbium doped fiber amplifier, OBPF: optical bandpass filter, PBS: polarization beam splitter, OSA: optical spectrum analyzer, PID: proportional integral derivative system, PD: photodetector, PW: power-meter. (b) Input pulse profile. (c) Linear resonances of the cavity.

Our fiber cavity is coherently-driven with a 1552.4 nm-wavelength cw laser (linewidth <1 kHz), intensity-modulated to generate 1.1-ns square pulses [see Fig. 1(b)]. These pulses have a repetition rate of 17.54 MHz, matching the free-spectral range of our cavity. Intensity modulation enables increased peak driving power levels, while also circumventing stimulated Brillouin scattering [2]. The driving pulses are amplified by means of an erbium-doped fiber amplifier (EDFA) before injection into the cavity. At each round-trip, the driving field is superimposed on the intracavity signal through a 90:10 coupler made of standard single mode fiber (SMF28). A 99:1 SMF tap-coupler is also included to extract a part of the intra-cavity field for analysis. The output signal is characterized both spectrally and temporally, using respectively an optical spectrum analyzer (OSA) and a real-time 50-GHz bandwidth oscilloscope coupled to 70-GHz fast photodetectors. Polarization management of this experiment is achieved by means of three polarization controllers (PC). First, the state of polarization (SOP) of the driving field is controlled (PC1) before the input coupler in order to predominantly excite one of the polarization modes of the fiber ring. A second PC mounted directly onto the spun fiber cavity (PC2) is used to finely tune the level of intra-cavity phase birefringence through local mechanical stress. This enables fine adjustments of the relative linear cavity round-trip phase shift $\delta_\perp$ between the two polarization modes of the cavity over the full $[-\pi, \pi]$ range; we measure this quantity by observing the linear cavity resonances while scanning the laser wavelength [see Fig. 1(c)]. Note that the laser wavelength can also be locked at a fixed absolute detuning $\delta_0$ from a cavity resonance; this is achieved with a feedback loop monitoring the intracavity power [8]. Finally, at the cavity output, a polarization beam splitter (PBS), preceded by a third polarization controller (PC3), is used to analyze individually the two polarization components of the intracavity field.

To model our experiment, we take advantage of the quasi-isotropic nature of our fiber, and describe propagation along the cavity with two coupled nonlinear Schrödinger equations. Denoting as $u^{(n)}(z,t)$ and $v^{(n)}(z,t)$ the circular polarization components of the intracavity electric field envelope during the $n^{th}$ round-trip [2], the field evolution obeys

$$\begin{cases} \partial_z u^{(n)} = -i\frac{\beta_2}{2}\partial_{tt}^2 u^{(n)} + i\gamma\left(\left|u^{(n)}\right|^2 + 2\left|v^{(n)}\right|^2\right)u^{(n)} \\ \partial_z v^{(n)} = -i\frac{\beta_2}{2}\partial_{tt}^2 v^{(n)} + i\gamma\left(\left|v^{(n)}\right|^2 + 2\left|u^{(n)}\right|^2\right)v^{(n)} \end{cases}. \quad (1)$$

Here, $z$ represents the propagation distance within the cavity, $t$ is time expressed in a delayed reference frame, and the nonlinearity coefficient $\gamma$ was taken to 4/W/km. To form a map, these equations are completed by boundary conditions to describe coherent superposition between the intra-cavity field and the driving field at each roundtrip, and also taking into account the lumped nature of our birefringence control. This is expressed in terms of the linear polarization components, $E_x = (u+v)/\sqrt{2}$ and $E_y = -i(u-v)/\sqrt{2}$, as:

$$\begin{cases} E_x^{(n+1)}(0,t) = (1-\alpha)E_x^{(n)}(L,t)e^{-i\delta_0} + \sqrt{\theta(1-\varepsilon)}E_x^{(in)} \\ E_y^{(n+1)}(0,t) = (1-\alpha)E_y^{(n)}(L,t)e^{-i(\delta_0+\delta_\perp)} + \sqrt{\theta\varepsilon}E_x^{(in)} \end{cases}. \quad (2)$$

Here the coefficient $\alpha = \pi/F = 0.116$ represents half of the total power loss per round-trip and $\theta = 0.1$ is the input coupling coefficient. Finally, for the sake of simplicity, we assume the driving field to be linearly polarized (along x) with a finite extinction ratio $\varepsilon$ ($\varepsilon \ll 1$) so as to account for realistic experimental conditions. This means that both circular components are almost evenly excited at the input of the cavity. Below, we refer to the driving power $P_{in} = \left|E_x^{(in)}\right|^2$ in terms of normalized quantity $X = \gamma L \theta P_{in}/\alpha^3$.

To highlight the key role of the birefringence in dissipative cavity PMI, we first discuss numerical results obtained by looking for steady-state solutions of the vector Ikeda map, Eqs. (1)–(2), for a range of relative detuning $\delta_\perp$. We use parameters matching those used in the experiment, namely a driving power $X = 28$ (corresponding to ~9 W peak driving power), a normalized linear detuning $\Delta = \delta_0/\alpha = 8$, and a finite polarization extinction-ratio between the linear components of the driving $\varepsilon = 10^{-3}$. The high driving power and detuning lead to high power PMI sidebands, hence enabling easier observations, but cavity PMI in fact occurs for a wide range of detunings. The pseudo-color plot in Fig. 2(a) shows optical spectra obtained from our simulations. We can clearly observe the presence of intra-cavity MI, with the generation of sidebands and several harmonics around the central driving frequency for a range of values of $\delta_\perp$. The strong dependence of the sidebands' position on the cavity birefringence clearly reveals the vector nature of the underlying process, and also that the sidebands are associated with PMI. We can notice that the plot exhibits an overall "umbrella shape," which repeats with a period of $\pi$ in $\delta_\perp$. No PMI is observed for $\delta_\perp$ very close to 0, i.e., for isotropic conditions, possibly due to the competing influence of the symmetry breaking instability that exists in such a case [20, 25, 26]. For other values of $\delta_\perp$, the PMI sidebands move away from the pump as $\delta_\perp$ increases, until they eventually disappear for $\delta_\perp$ mod $\pi > 0.8\pi$. We attribute this extinction to a decrease in the PMI gain.

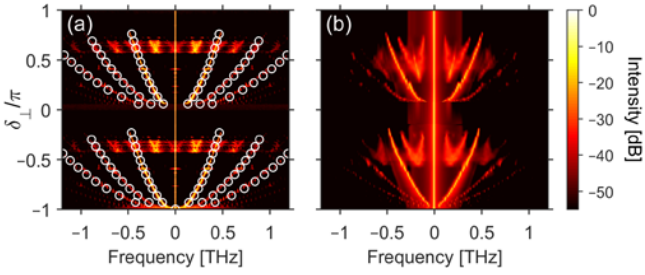

**Fig. 2**. Pseudo-color plots of (a) numerical and (b) experimental output spectra for various values of relative detuning $\delta_\perp$, i.e., intra-cavity birefringence. Circles in (a) highlight how the position of the PMI sidebands match the theoretical predictions of Eq. (4). Driving power $X = 28$ and linear detuning $\Delta = 8$. Spectra are plotted as a function of frequency detuning from the central cw driving component.

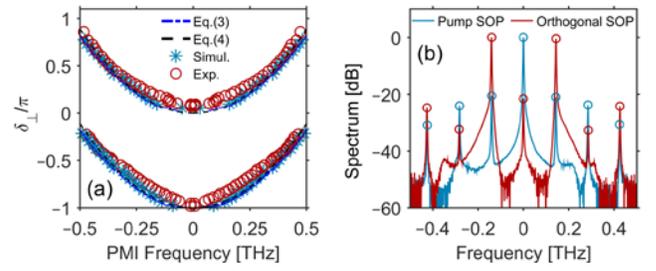

**Fig. 3**. (a) Optimum frequency ($\Omega_{opt}$) of the first-order PMI sidebands versus relative detuning $\delta_\perp$. Experimental results (red circles) are compared with numerical simulations (blue stars) as well as theoretical predictions of Eqs. (3)–(4) (blue and black dashed-lines, respectively). (b) Output spectrum for $\delta_\perp = 0.2\pi$ recorded after the output PBS and decomposed into polarization components parallel (blue) and orthogonal (red) to the driving SOP.

The optimum angular frequency of the PMI process, $\Omega_{opt}$, can be approximated by the following phase-matching condition, that expresses a balance between chromatic dispersion, linear detuning, birefringence, and intra-cavity power ($P$):

$$\Omega_{opt}^2 \frac{\beta_2 L}{2} - (\delta_\perp \bmod \pi + \delta_0) + \gamma P L = 0 \quad (3)$$

Note that, for $\delta_\perp$ close to zero, this phase-matching condition corresponds to the isotropic PMI discussed in Ref. [20]. Since $\gamma P L - \delta_0$ is relatively small in Eq. (3), the expression for the optimum PMI frequency can be further simplified as:

$$\Omega_{opt} = \sqrt{\frac{2[\delta_\perp \bmod \pi]}{\beta_2 L}} \quad (4)$$

The expression above, represented with circles in Fig. 2(a), clearly highlights the role of birefringence in this process.

In Fig. 2(b), we show a concatenation of experimental spectra measured at the output port of the 99:1 tap coupler in the same conditions as the numerical results of Fig. 2(a). The relative detuning $\delta_\perp$ was adjusted step-by-step by means of the intra-cavity PC2 while the detuning was kept locked at $\Delta = 8$. One can observe a good qualitative agreement between simulations and experiments. In particular, the repeating "umbrella" shape is clearly visible, as are the well separated, far detuned, and narrow sidebands corresponding to PMI. We must note an additional region of parametric gain around $\Omega_{opt}/2$, present in both the numerical and experimental spectra. We attribute this phenomenon to a more complex vectorial four-wave mixing process occurring between the pump wave and the first elliptically-polarized sidebands. Figure 3(a) presents a further comparison of the position of the first-order PMI sidebands obtained numerically (stars), analytically [dashed curves, Eqs. (3) and (4)], and experimentally (circles) as a function of the relative detuning $\delta_\perp$. The agreement is very good.

To assess clearly the vector nature of the observed PMI sidebands, we have characterized the output spectrum obtained for a relative detuning $\delta_\perp = 0.2\pi$ in terms of polarization components that are parallel (blue curve) and orthogonal (red curve) to the driving field's SOP, respectively [Fig. 3(b)]. The results clearly highlight that the first order PMI sidebands are mainly orthogonally polarized to the driving field, with an extinction-ratio better than 20 dB. Higher-order sidebands exhibit alternating SOPs as they result from cascaded PMI. This behavior is well-known in conventional isotropic PMI experiments performed in single-pass fiber segments [2, 19].

In addition to the spectral analysis shown in Figs. 2 and 3, we have also performed a temporal characterization of the intra-cavity intensity profile in the PMI regime using a 50-GHz bandwidth real-time oscilloscope. An example of measurement taken over a single cavity round-trip is shown in Fig. 4(a), where we can observe temporal oscillations at a frequency of about 15 GHz across our ns-long driving pulses. Here we have selected a value of $\delta_\perp$ slightly positive, and low enough for the PMI frequency to fit within the bandwidth of our oscilloscope and photodetectors. Also, observations are made in a polarization basis at 45° with respect to the SOP of the driving field and the PMI sidebands so as to mix the cw driving component with the PMI signal. In this way, the observed oscillations in intensity seen in both orthogonal polarization components (blue and red curves) are essentially proportional to the PMI sideband field *amplitude*. This provides maximal contrast and reveals oscillations at the actual PMI frequency [rather than its second harmonic, as $\cos^2(\Omega_{opt}t) \propto 1+\cos(2\Omega_{opt}t)$]. The oscillations are also anti-correlated across the two components, which is typical of vector MI processes [17–20].

Additional insights into the temporal dynamics of intracavity PMI can be gained by monitoring the temporal intensity profile of the generated patterns round-trip by round-trip. An example of such measurement is shown in Fig. 4(b) in the form of a pseudo-color plot, showing data for 20 successive round-trips (bottom to top). The data was acquired single-shot by our oscilloscope as a long temporal sequence, which was then split into individual cavity round-trips. Here we only show a single polarization component, in the same basis as in Fig. 4(a), and $\delta_\perp$ is again small and positive. We can observe that the PMI pattern (here with a frequency of 27.5 GHz) is stable and repeats identically from round-trip to round-trip. In Fig. 4(c), we have also plotted the corresponding total intensity measured independently. This plot reveals hints of oscillations at twice the PMI frequency. Together with the low contrast (which is partly limited by the bandwidth of our oscilloscope), this further confirms the vector nature of the observed patterns. Figures 4(d)–(e) display the results of a similar measurement but obtained for a relative detuning close to $-\pi$, i.e., close to anti-resonant conditions. In this case, as can be appreciated from the checkerboard-like pattern, the PMI pattern flips at every cavity round-trip, in a way reminiscent to a period-doubled dynamics (P2) [27]. This behavior can be understood by noting that a relative detuning $\delta_\perp \sim \pm\pi$ makes the system acts as a half-wave plate, swapping the handedness of the circular polarization components at each round-trip, and leading to the observed flipping [22, 28]. Interestingly, the total intensity [Fig. 4(e)] shows no sign of the periodic flipping, confirming that this behavior stems from a pure polarization dynamics.

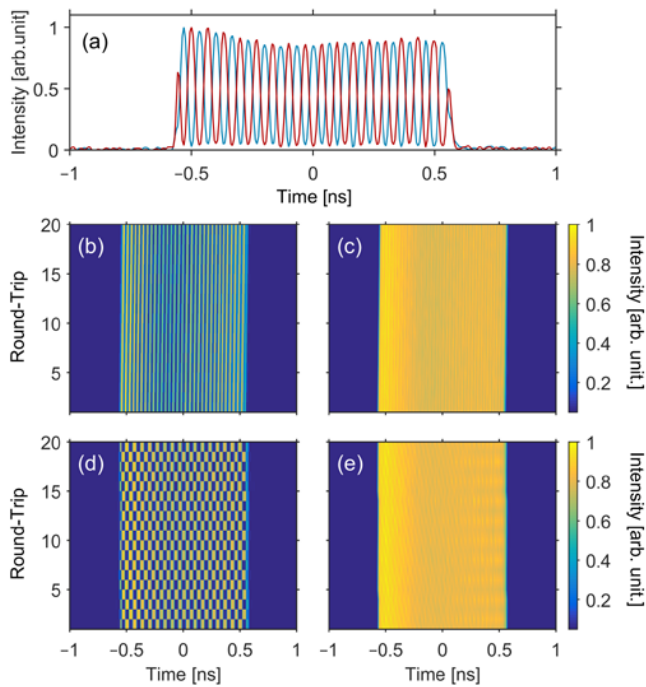

**Fig. 4**. (a) Typical experimentally measured temporal intensity profiles of both orthogonal polarization components (at 45° with respect to the driving and PMI sidebands SOPs) monitored at the output of the cavity for $\delta_\perp$ slightly positive (low PMI frequency, within the bandwidth of our detectors). (b) Round-trip by round-trip evolution (from bottom to top) of the temporal intensity profile at the output of our fiber cavity monitored on one of the polarization components (as used in panel (a)) for $\delta_\perp$ slightly above 0. (c) Corresponding total intensity (before the PBS). (d, e) Same as (b, c) but with $\delta_\perp$ slightly above $-\pi$.

In conclusion, we have reported the experimental observation of polarization modulation instability in a fiber Kerr resonator. It was shown that the phase matched PMI frequency is affected by the level of intra-cavity birefringence and can be well approximated with a simple equation, Eq. (4). Our experimental observations were performed in a 12-m long normal dispersion spun fiber ring cavity incorporating a polarization controller for accurate control of the level of birefringence. By means of real-time temporal characterizations, both steady and period-doubled (P2) behaviors have been observed. Experimental results were found to be in good agreement with theoretical predictions and numerical simulations based on a two-component Ikeda map. These results establish birefringence as a potential new degree of freedom for frequency comb generation in normal dispersion Kerr resonators, and provide new insights into the control of dissipative MI.

**Funding.** JF and BK acknowledge the Conseil Régional de Bourgogne Franche-Comté, FEDER and ANR (ANR-15-IDEX-03). We acknowledge The Royal Society of New Zealand, Marsden Funding (18-UOA-310), and James Cook (JCF-UOA1701, for SC) and Rutherford Discovery (RDF-15-UOA-015, for ME) Fellowships.